# Cost-efficient, QoS and Security aware Placement of Smart Farming IoT Applications in Cloud-Fog Infrastructure

Jagruti Sahoo
*Dept. of Comp. Science and Mathematics,*
*South Carolina State University*
Orangeburg, USA
jsahoo@scsu.edu

*Abstract*— **Smart farming is a recent innovation in the agriculture sector that can improve the agricultural yield by using smarter, automated, and data driven farm processes that interact with IoT devices deployed on farms. A cloud-fog infrastructure provides an effective platform to execute IoT applications. While fog computing satisfies the real-time processing need of delay-sensitive IoT services by bringing virtualized services closer to the IoT devices, cloud computing allows execution of applications with higher computational requirements. The deployment of IoT applications is a critical challenge as cloud and fog nodes vary in terms of their resource availability and use different cost models. Moreover, diversity in resource, quality of service (QoS) and security requirements of IoT applications make the problem even more complex. In this paper, we model IoT application placement as an optimization problem that aims at minimizing the cost while satisfying the QoS and security constraints. The problem is formulated using Integer Linear Programming (ILP). The ILP model is evaluated for a small-scale scenario. The evaluation shows the impact of QoS and security requirement on the cost. We also study the impact of relaxing security constraint on the placement decision.**

I. INTRODUCTION

Smart Farming has brought a significant revolution to the agriculture field by allowing farmers to perform farm operations with greater efficiency and make informed decisions [1][2]. It also increases the agricultural yield which is critical to address the global food demand in next few years. A smart farming system consists of IoT nodes (e.g., soil sensors, pH probes, temperature/humidity sensor, etc.,) deployed on a farm and provides the basis for applications including soil monitoring, crop monitoring, and precision agriculture [3].

Typically, an IoT-based smart farming system uses a remote cloud server for storage and computation. However, latency-sensitive IoT services involve real-time data processing which is difficult to achieve with a cloud server because of higher communication latency. Fog computing is an effective paradigm that reduces the latency by provisioning virtualized computational, storage, and networking resources closer to the edge where the data is consumed [4]. Since, the fog nodes have limited capacity, delay-tolerant services with higher resource requirement can be deployed on the cloud server. A hybrid infrastructure consisting of both cloud and fog nodes is suitable for building IoT systems, to leverage the benefits of both paradigms. Fig. 1 shows a high-level view of a smart farming system built using cloud and fog computing. The fog layer includes wireless network of fog nodes deployed in a farm area. We consider that smart farming IoT applications are developed using distributed data flow (DDF) model, where each application is composed of interdependent application modules, each with a specific resource (e.g. CPU, memory and storage) requirement [5][6]. Each IoT application is associated with a QoS value that indicates the delay threshold. The fog nodes have limited capacity and demands more cost over cloud for using same amount of resources. As a result, random placement of application modules in the cloud-fog infrastructure will affect the cost and may not satisfy the heterogeneous QoS requirement. Therefore, it is a critical challenge to determine the optimal placement of IoT application modules that minimize cost and provide a QoS guarantee.

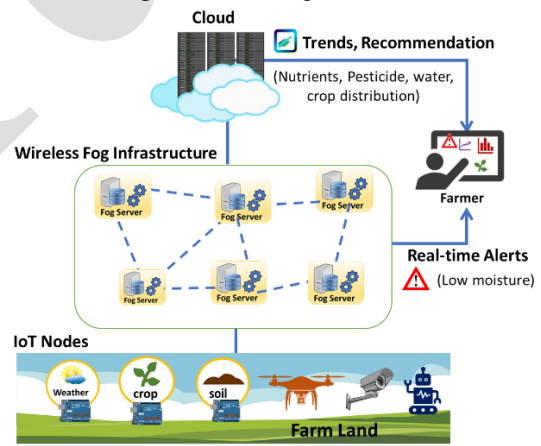

Fig. 1. Smart Farming System

Data security is an important requirement that has not been studied before in making the placement decisions. The massive amount of data generated by smart farming sensors make it an attractive target for data theft. Typically, fog nodes offer more security over cloud because of limited Internet use [7]. However, the broadcast nature of wireless communications makes some fog nodes vulnerable to sniffing. Even if physical intrusion is not possible, attackers can still sniff farm data as the fog nodes have very high transmission range that extends beyond the farm area. Although the farm data may be encrypted, with advanced tools, attackers can decrypt the key and compromise sensitive farm variables and disclose the data for financial gain. Because of the security disparity of the cloud and fog nodes, it is imperative to select the resource nodes in a way that provides the required protection to the farm data.

In this paper, we address the IoT application placement in a hybrid cloud-fog infrastructure and modeled it as an optimization problem using ILP. The problem involves minimizing the resource cost while satisfying the QoS and security requirement of IoT applications. We study the optimal solution in a small-scale setting and observe the objective function under different QoS scenarios. We also observe the impact of relaxing security constraint on the placement decision.

The remainder of the paper is organized as follows. Section. II discusses the related works. The ILP of the IoT application placement problem is presented in Section. III. Section. IV presents the simulation results. Section. V provides the conclusion and future work.

## II. Related Works

IoT application placement has been studied in the literature to find a suitable location in a cloud-fog infrastructure to execute the IoT applications that process the data received from the IoT nodes [8]. Optimization models have been proposed to address the placement problem that includes a variety of objectives including cost [9][10], QoS [11], Quality of Experience (QoE) [12], resource utilization [13][14], and energy consumption [15]. Yousefpour et al. [9] proposed a resource provisioning scheme which involves dynamically deploying new applications on fog and cloud servers in a way that minimizes the resource cost as well as the cost of delay violations. The threshold on desired service delay of the IoT applications was used to model the QoS constraint.

Mai et al. [11] proposed a task assignment approach in which IoT tasks are assigned to appropriate fog servers to minimize the QoS (i.e., computational latency that consists of propagation, execution and buffering related latency). Mai et al. used a reinforcement learning based algorithm in which a softmax action selection function is used to select the server. The task assignment seems to be a variation of the IoT application placement. Although, the algorithm by Mai et al. [11] satisfies the QoS requirement, it lacks cost-efficiency.

Mahmud et al. [12] proposed a QoE-aware application placement that aims at placing the IoT application based on user expectations and the current capabilities of fog instances such as round-trip time, resource availability and processing speed. Mahmud et al. [12] considers an application model in which an IoT application is divided into two modules: client module and main module. The main module is deployed on a fog server whereas the client module runs at the end-user device (e.g., smartphone). The main module contains operations such as data filtration, data analysis and event processing. It communicates with the client module and collects the user expectations and the results needed by the end-user. Although the work by Mahmud et al. [12] provides enhanced user QoE, it lacks cost-efficiency as it ignores the resource cost.

Skarlat et al. [13] proposed a resource-aware placement in a fog infrastructure that consists of a hierarchy of fog colonies with the top most fog colony residing in cloud. The fog colony are micro data centers and consists of fog cells which provide the virtualized resources for task execution and are responsible for coordinating IoT devices. Skarlat et al. [13] aims at maximizing the number of services that can run in the fog infrastructure while minimizing delay that may incur by assigning tasks to a higher-level fog colony. Minh et al. [14] also maximizes the number of services that are deployed in the fog and provides a QoS guarantee. Although The placement provides a QoS guarantee Although, resource-aware placement schemes ensure efficient utilization of the fog resources, they do not necessarily reduce the cost associated with the placement.

Goudarzi et al. [15] discussed an optimization problem that jointly minimizes the execution time and energy consumption to place the applications of multiple IoT devices. The execution time includes offloading latency of tasks of an application, computing time, and data transmission time, thereby optimizing the QoS. Memetic algorithm is used Goudarzi et al. [15] to design a batch application placement scheme that maps the tasks belonging to concurrent IoT applications on appropriate cloud or fog servers.

Unlike the above works, we consider the security requirement of IoT applications in placing them on appropriate resource nodes. There have been a handful of work on security aware IoT application placement [16][17]. Auluck et al. [16] addresses the scheduling of IoT applications among fog and cloud data center considering the deadline and security constraints of the applications. The deployment of an application is restricted to a specific resource node (local fog, remote fog, private cloud, and public cloud) based on the privacy-level of the application. The main issue with [17] is that it does not distinguish between local fog nodes in terms of their security status. Not all fog nodes can provide identical protection due to their heterogeneous system configuration and services. We, therefore, design a security constraint that takes into consideration fog nodes with different security status.

## III. IoT Application Placement

### A. Application Model

We consider a linear model for representing the smart farming applications. Each application consists of a linear chain of modules, each with specific requirement for processing, memory, and storage resources. Each module processes the data received from a previous module in the linear chain and produces output that is consumed by the next module in the chain. Fig. 1 shows a crop monitoring application that consists of three modules: 1) *sense*: that collects data from IoT nodes (i.e., crop sensors) and removes noisy data, 2) data aggregation: that aggregates sensory information from multiple sensors and applies temporal or spatial aggregation to remove redundant data and improve the data quality, 3) alerts: that analyzed the aggregated data and provide farmers with timely alerts on damaged crops.

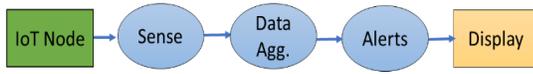

Fig. 2. Crop Monitoring Application

### B. Problem Statement

We aim to address the IoT application placement problem that entails finding the optimal resource nodes in a cloud-fog infrastructure for executing a number of smart farming applications in a way that minimizes the resource cost. Moreover, the placement must satisfy the resource needs, QoS and security requirement of the smart farming applications.

### C. ILP Model

In this section, we present an ILP formulation of the IoT application placement problem. The notations used in the formulation are provided in Table I. Our objective is to minimize the resource cost. The resource cost has three components: processing cost, storage cost, and communication cost. The objective is modelled as follows:

Minimize:

$$\sum_{i \in A} \sum_{j \in A_i} \sum_{k \in N} T_{ij} \, c_k^{proc} \, x_{ijk}$$
$$+ \sum_{i \in A} \sum_{j \in A_i} \sum_{k \in N} R_{ij}^{stor} \, c_k^{stor} \, x_{ijk}$$
$$+ \sum_{i \in A} \sum_{u \in N} \delta^{sensor} \, c_u^{sensor} x_{i1u}$$
$$+ \sum_{i \in A} \sum_{j \in A_i} \sum_{u \in N} \sum_{v \in N} \delta_{ij} \, c_{uv}^{bw} \, z_{ijuv}$$
$$+ \sum_{i \in A} \sum_{u \in N} \delta^{user} \, c_u^{user} x_{inu} \quad (1)$$

Where, $x_{ijk}$ is a binary decision variable that determines whether module $j$ of application $i$ is placed on the resource node $k$. The first and second term represents the processing and storage cost, respectively. The third term represents the communication cost incurred when a sensor node transmits its data to a resource node $u$. It includes the variable $x_{i1k}$ as the data from sensor node is received by module 1 of the application. The fourth term, represents the communication cost of transmitting data between two resource nodes $u$ and $v$. The last term represents the communication cost of transmitting the results to the end-user. It includes the variable $x_{ink}$ as the last ($n^{th}$) module of the application transmits data to the end-user.

*1) Resource Constraints:* Each resource node has a certain processing, memory, and storage capacity. This constraint ensures that the resource requirements of application modules placed on a resource node must not exceed the capacity of the node, and is formulated as follows:

$$\sum_{i \in A} \sum_{j \in A_i} R_{ij}^{proc} \, x_{ijk} < P_k, \forall k \in N \quad (2)$$

$$\sum_{i \in A} \sum_{j \in A_i} R_{ij}^{mem} \, x_{ijk} < M_k, \forall k \in N \quad (3)$$

$$\sum_{i \in A} \sum_{j \in A_i} R_{ij}^{stor} \, x_{ijk} < S_k, \forall k \in N \quad (4)$$

*2) QoS Constraint:* We consider that each application is associated with a QoS requirement that represents the maximum end-to-end delay the application can tolerate. End-to-end delay is the total time period from the moment data is received from sensors till the processed results are delivered to the end-user. End-to-end delay consists of two components, communication delay and execution delay. Communication delay involves the delay of communication between a sensor and the first application module, inter-module communication, and the communication between the last module and the end-user. Execution delay is the delay of executing the modules on the resource nodes. The communication delay, $D_{comm}$ and execution delay, $D_{exec}$ are obtained as follows:

$$D_{comm} = \sum_{i \in A} \sum_{u \in N} d_u^{sensor} x_{i1u}$$
$$+ \sum_{i \in A} \sum_{j \in A_i} \sum_{u \in N} \sum_{v \in N} t_{uv} \, z_{ijuv}$$
$$+ \sum_{i \in A} \sum_{u \in N} d_u^{end-user} x_{inu} \quad (5)$$

$$D_{exec} = \sum_{i \in A} \sum_{j \in A_i} \sum_{k \in N} T_{ij} \, x_{ijk} \quad (6)$$

The QoS constraint ensures that the end-to-end delay must not exceed the delay threshold of IoT applications and is expressed as follows:

$$D_{comm} + D_{exec} < Q_i, \forall i \in A \quad (7)$$

TABLE I. ILP NOTATIONS

| Notation | Meaning |
|---|---|
| $N$ | Set of resource nodes |
| $A$ | Set of applications |
| $A_i$ | Set of modules that constitute application $i$ |
| $n$ | Number of application modules |
| $c_k^{proc}$ | Processing cost per sec at node $k$ |
| $c_k^{stor}$ | Storage cost per Gb per sec at node $k$ |
| $c_{uv}^{bw}$ | Bandwidth cost per Gb per sec of the physical link $(u, v)$ |
| $c_u^{sensor}$ | Bandwidth cost per Gb per sec of the physical link between sensor and node $u$ |
| $c_u^{user}$ | Bandwidth cost per Gb per sec of the physical link between user and node $u$ |
| $P_k$ | Processing Capacity of node $k$ (MIPS) |
| $M_k$ | Memory Capacity of node $k$ (Gb) |
| $S_k$ | Storage Capacity of node $k$ (Gb) |
| $R_{ij}^{proc}$ | Processing Requirement of module $j$ of application $i$ (MI) |
| $R_{ij}^{mem}$ | Memory Requirement of module $j$ of application $i$ (MI) |
| $R_{ij}^{stor}$ | Storage Requirement of module $j$ of application $i$ (Gb) |
| $\delta_{ij}$ | Size of traffic exchanged between module $i$ and $j$ (Gb) |
| $d_u^{sensor}$ | Communication delay between a sensor and resource node $u$ |
| $d_u^{end-user}$ | Communication delay between a resource node $u$ and end-user |
| $t_{uv}$ | Communication delay between resource node $u$ and $v$ |
| $T_{ij}$ | Execution delay of module $j$ of application $i$ |
| $W_i$ | Security requirement of application $i$ |
| $\rho_k$ | Security rating of resource node $k$ |
| $Q_i$ | QoS Threshold of application $i$ |
| **Parameter** | **Variable** |
| $x_{ijk}$ | 1 if module $j$ of application $i$ is deployed on resource node $k$, 0 otherwise |
| $z_{ijuv}$ | 1 if the application edge $(j, j+1)$ of application $i$ is mapped to the physical edge $(u, v)$, 0 otherwise |

*3) Security Constraint:* We consider that some of the smart farming applications process sensitive data and hence the application modules need to be placed on a resource node that provides maximum protection from unwanted disclosures and data breaches. Each application $i$ is associated with a security requirement, $W_i$ that represents the sensitivity of the data processed by the application. $W_i$ also represents the minimum-security requirement that a resource node must have for executing the application modules. $W_i$ is expressed using one of the three levels: low, medium, and high represented numerically by 1, 2, and 3, respectively. Each resource node $k$ is associated with a security rating, $\rho_k$ that shows the ability of the node to protect data from unwanted disclosures. We consider the same three levels for representing the security rating. We consider a wireless fog infrastructure. As a result of the wireless communication, attackers may attempt to capture sensitive data using sniffing attacks. This is because, the transmission range of the fog nodes may extend well beyond the farm area. We assign a rating of "low" to fog nodes that are prone to sniffing attacks. Other fog nodes are assigned a rating of "high". Cloud server is assigned a "medium" rating as the data will travel through the Internet before it is received by the destination application module, thereby increasing the risk of data breaches. However, cloud server is preferred over the fog nodes that are prone to sniffing attack as wired links are more secure than wireless links.

To determine whether a fog node is prone to sniffing attack, we will compute the distance between the fog node's position and all four boundaries of the farm area. If any of the distance values is less than $R$ (transmission range), then the transmission range extends beyond the farm boundary, and hence allows attackers to capture the communication. Fig. 2 shows two fog nodes $F_1$ and $F_2$ located in a farm. The dotted circle shows the transmission range. Since $D_2 < R$, $F_2$ is prone to sniffing attack, whereas $F_1$ is secure from sniffing attacks. The shaded area shows the sniffing area within the transmission range of fog node $F_2$.

We design the security constraint as follows:
$$\sum_{k \in N} \rho_k \, x_{ijk} \geq W_i , \forall j \in A_i, \forall i \in A \quad (8)$$

It ensures that the security rating, $\rho_k$ of a resource node $k$ that hosts a module $j$ of application $i$ must be greater than the security requirement of application $i$ denoted as $W_i$.

*4) Others:* Next, we formulate a constraint to ensure that each application module is placed on exactly one resource node. This constraint is given by (9).
$$\sum_{k \in N} x_{ijk} = 1 \; \forall i \in A, \forall j \in A_i \quad (9)$$

In addition to mapping the modules to resource nodes, we also map the logical links to the physical links. The following constraint ensures that the logical link $(j, j')$ is mapped to the physical link $(u, v)$ only if the resource nodes $u$ and $v$ host the application modules $j$, and $j'$, respectively.

$$x_{ijk} * x_{ij'k} = z_{ijuv}, \forall i \in A, j \in A_i, j' \in A_i,$$
$$j' = j + 1, u \in N, v \in N \quad (10)$$

(10) is a quadratic constraint and can be linearized as follows:
$$z_{ijuv} \leq x_{ijk}, \forall i \in A, j \in A_i, j' \in A_i, j' = j + 1,$$
$$u \in N, v \in N \quad (11)$$
$$z_{ijuv} \leq x_{ij'k}, \forall i \in A, j \in A_i, j' \in A_i, j' = j + 1,$$
$$u \in N, v \in N \quad (12)$$
$$z_{ijuv} \geq x_{ijk} + x_{ij'k} - 1, \forall i \in A, j \in A_i, j' \in A_i,$$
$$j' = j + 1, u \in N, v \in N \quad (13)$$

We also formulate a constraint to ensure that each logical link is mapped to exactly one physical link. The constraint is given below.
$$\sum_{u \in N} \sum_{v \in N} z_{ijuv} = 1 \; \forall i \in A, \forall j \in A_i \quad (14)$$

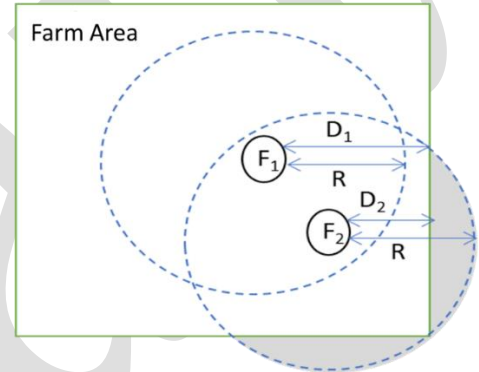

Fig. 3. Finding Security Rating of Fog Node.

## IV. PERFORMANCE EVALUATION

### A. Simulation Set up

We use IBM Cplex optimization studio [18] to implement the proposed ILP model. We consider a hybrid cloud-fog infrastructure consisting of a cloud server and two fog nodes. Each IoT application consists of three modules. For each module, the processing (in MI), memory (in Gb), and storage (in Gb) requirements are generated at random in the interval (100, 2100)X $10^{-3}$, (10, 40) X $10^{-3}$, and (256, 768) X $10^{-3}$, respectively. For each application, the QoS value is generated at random in the interval (MinQoS, MaxQoS), where MinQoS is fixed at 0.5s, whereas MaxQoS takes two values: 1.5s and 3s, resulting in different simulation scenarios. The security level of each application is generated at random from {1, 2, 3}, where "low", "medium", and "high" levels are indicated by the numeric values: 1, 2, and 3, respectively. The simulation parameters are listed in Table. II.

### B. Performance Metrics

We consider the following metrics to evaluate the effectiveness of the proposed algorithm.

*1) Resource Cost:* It is defined as the total cost of processing, storage, and communication resources required to

execute the IoT applications in the hybrid cloud-fog infrastructure.

*2) Number of Deployed Modules:* It is the number of modules deployed on each type of resource: cloud and fog.

*3) Amount of Unprotected Data:* It is defined as the amount of data that is prone to disclosures when a resource node that fails to provide the required protection. This metric is obtained by relaxing the security constraint in the proposed ILP model.

Table. II Infrastructure Parameters

| Parameter | Value |
|---|---|
| Number of fog nodes | 2 |
| Number of Applications | 1-7 |
| Number of modules per application | 3 |
| Size of Input traffic (Gbps) | 1.0- 4.0 (x $10^{-3}$) |
| Size of traffic exchanged between adjacent modules (Gbps) | 0.1-1.0 |
| Size of Output Traffic (Gbps) | 0.5-1.0 (x $10^{-3}$) |
| Processing Cost in Cloud per sec | 0.03 |
| Processing Cost in fog node per sec | 0.02 |
| Storage Cost in Cloud per Gb per sec | 0.001 |
| Storage Cost in fog per Gb per sec | 0.02 |
| Communication Cost (Cloud) per Gb per sec | 3.0 |
| Communication Cost (Fog) per Gb per sec | 5.0 |
| Communication Delay (Cloud-Fog) | 500msec |
| Communication Delay (Fog-Fog) | 10ms |

*C. Results & Discussions*

Fig. 4 shows the resource cost with respect to the number of applications under two different QoS scenarios indicated by MaxQoS. We observe a steady increase in resource cost as the number of applications increases irrespective of the QoS scenario. The resource cost with a MaxQoS of 1.5s is higher than that with a MaxQoS of 3s. This is because when MaxQoS is lower, more modules are placed on the fog nodes than cloud because of the stringent QoS requirement. Moreover, fog resources are expensive than the cloud resources, resulting in a higher cost. With a MaxQoS of 3s, most of the applications have a higher QoS requirement, and hence are placed in the cloud leading to a lower cost.

Fig. 5 shows the cost for the same two QoS scenarios with respect to the fraction of applications that have "high" security requirement, represented as α. Note that modules acquire the same security requirement as the application that is made up of them. As α increases, we observe an increase in cost. The modules with "high" security requirement can only be placed on the fog node with "high" security rating. Since fog resources cost more compared their cloud counterpart, the more the modules with "high" security requirement, more is the cost to place them. Moreover, out of two QoS scenarios, the one with the higher value of MaxQoS threshold shows a slightly lower cost than the other scenario. This cost savings is achieved because higher QoS scenario involves more applications with either "low" or "medium" security requirement compared to the lower QoS scenario. As a result, modules of those applications are placed on the cloud, resulting in a lower cost.

Fig. 6 shows the number of modules placed on cloud and fog with respect to the number of applications under two different QoS scenarios. When the number of applications increases, fog continues to hosts more modules than cloud. For 7 applications, fog hosts twice as much as the cloud when MaxQoS is 1.5s. For the same QoS scenario, we observe a 600% increase in modules placed on the fog. We observe that MaxQoS of 1.5s results in more fog modules than when MaxQoS is 3s. This is because MaxQoS of 1.5s results in majority of the applications having stringent QoS requirement, thereby preventing the constituent modules from being deployed on the cloud. On the other hand, the placement on cloud shows a contrasting behavior. The scenario with MaxQoS of 3s shows deployment of more modules on cloud compared to when MaxQos is 1.5s. This is due to the increase in number of applications with less stringent QoS requirement when MaxQoS is 3s, resulting in more modules being placed on the cloud.

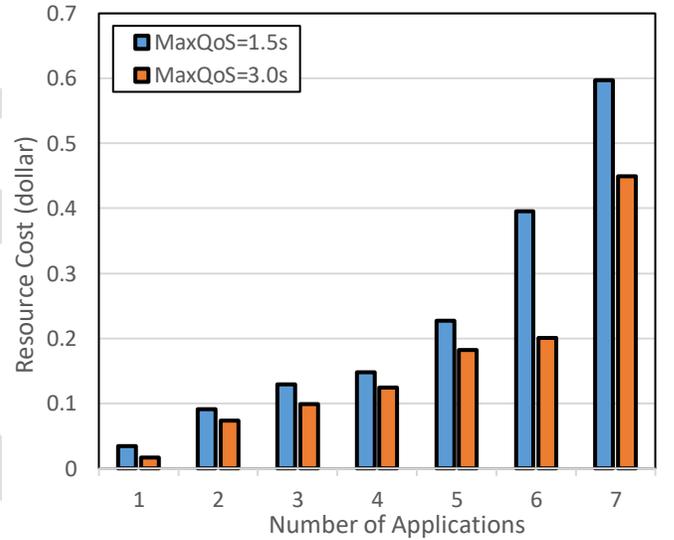

Fig. 4. Resource Cost

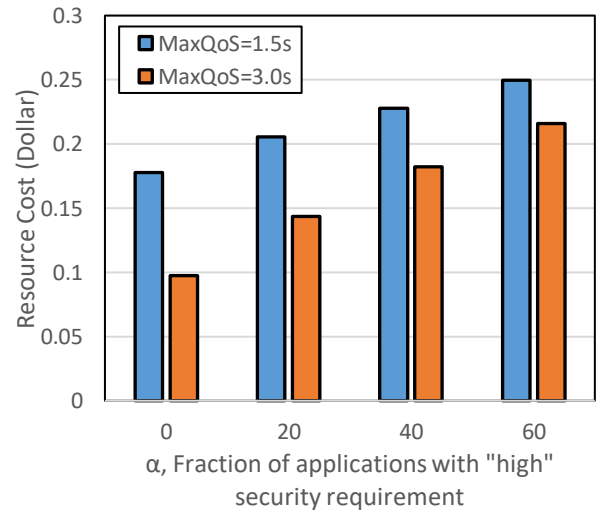

Fig. 5. Resource Cost (Security requirement)

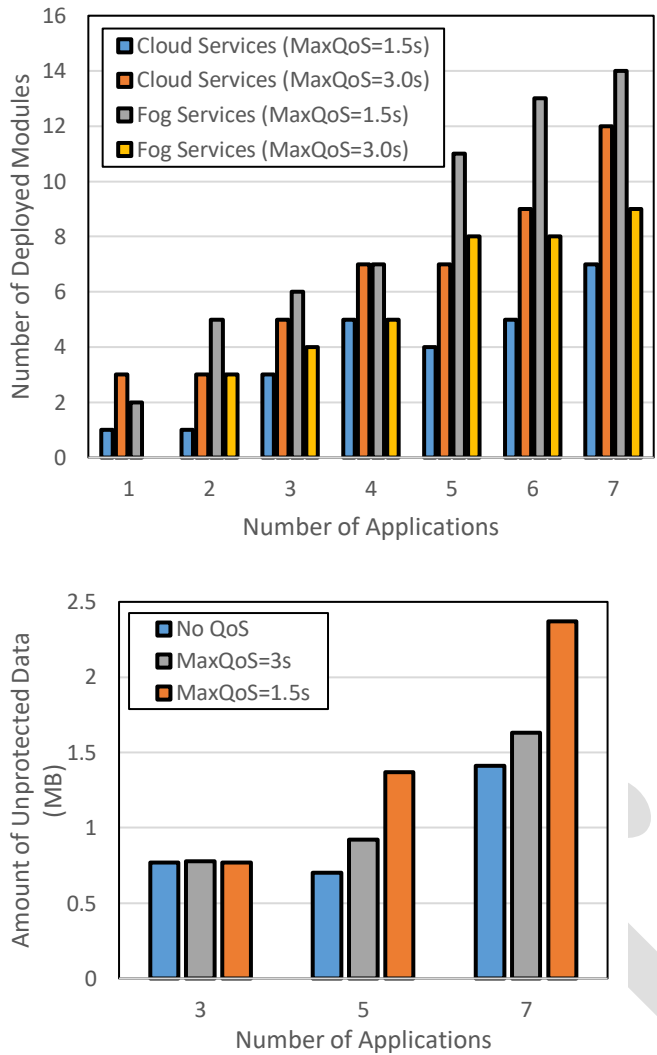

higher amount unprotected data than in NoQoS scenario. This is because, those types of applications are placed on the fog nodes due to their QoS requirements. As one of the fog nodes have "low" rating, it results in security violations. Between the two QoS scenarios, the one with of 1.5s involves higher number of applications with either "medium" or "high" security requirement and hence more prone to security consequences. The above observations show the importance of incorporating security constraint to ensure the data security.

## V. CONCLUSION & FUTURE WORK

In this paper, we address the IoT application placement problem for a hybrid cloud-fog based smart farming system. The placement entails minimizing cost while ensuring QoS of the applications and security of data processed, stored, or transmitted by the applications. The problem is formulated as using ILP. We evaluate the optimal solution in a small-scale scenario and study the cost of placing applications in different QoS scenarios. We also study the need to incorporate security requirement in application placement in order to ensure that the farm data remain protected from unwanted disclosures. In future, we will be interested in studying the application placement considering a hierarchical fog network. We would like to extend the proposed ILP to incorporate multiple risk scenarios in designing the security constraint.

## ACKNOWLEDGMENT

This work is supported by the National Institute of Food and Agriculture, United States Department of Agriculture, Evans-Allen project number SCX-314-02-19.

Fig. 7. Amount of Unprotected Data

Fig. 7 shows amount of unprotected data that with respect to the number of applications in three different scenarios: NoQoS, MaxQoS=1.5s, and MaxQoS=3.0s with α set to 0.25 (i.e., 25% of the applications have "high" security requirement). Note that in NoQoS scenario, both QoS and security constraints are relaxed; whereas other two QoS scenarios involves a relaxation of security constraint only. All three scenarios show the same amount of unprotected data when there are 3 applications. The amount of unprotected data increases when the number of applications increases to 7. The NoQoS scenario shows the least amount of unprotected data, whereas the highest amount (2.37Mb) attributes to the scenario with a MaxQoS of 1.5s. As the cost needs to be minimized, in NoQoS scenario, all applications are placed on the cloud. Since cloud has a "medium" rating, no security violations occur for applications with either "low" or "medium" security requirement. Only applications with "high" security requirement experience security violation. However, when QoS constraint is introduced, applications with either "medium" or "high" security requirement experience security violations resulting in

## REFERENCES

[1] O. Elijah, T. A. Rahman, I. Orikumhi, C. Y. Leow and M. N. Hindia, "An Overview of Internet of Things (IoT) and Data Analytics in Agriculture: Benefits and Challenges," in *IEEE Internet of Things Journal*, vol. 5, no. 5, pp. 3758-3773, Oct. 2018.

[2] C. Brewster, I. Roussaki, N. Kalatzis, K. Doolin and K. Ellis, "IoT in Agriculture: Designing a Europe-Wide Large-Scale Pilot," in *IEEE Communications Magazine*, vol. 55, no. 9, pp. 26-33, Sept. 2017.

[3] E. Navarro, N. Costa, A. Pereira, "A Systematic Review of IoT Solutions for Smart Farming", in *Sensors*. 2020, vol. 20, no. 15, 4231.

[4] C. Mouradian, D. Naboulsi, S. Yangui, R. H. Glitho, M. J. Morrow and P. A. Polakos, "A Comprehensive Survey on Fog Computing: State-ofthe-Art and Research Challenges," in *IEEE Communications Surveys & Tutorials*, vol. 20, no. 1, pp. 416-464, Firstquarter 2018.

[5] N. K. Giang, M. Blackstock, R. Lea, and V. C. M Leung, "Distributed Data Flow: a Programming Model for the Crowdsourced Internet of Things", in *Proc. ACM Middleware Doct Symposium 2015*, New York, NY, USA, Article 4, 1–4.

[6] M. Taneja and A. Davy, "Resource aware placement of IoT application modules in Fog-Cloud Computing Paradigm, " in *Proc. 2017 IFIP/IEEE Symposium on Integrated Network and Service Management (IM)*, Lisbon, 2017, pp. 1222-1228.

[7] N. Tariq, M. Asim, F. Al-Obeidat, M. Zubair Farooqi, T. Baker, M. Hammoudeh, and I. Ghafir, "The Security of Big Data in Fog Enabled IoT Applications Including Blockchain: A Survey," in *Sensors,* vol. 19, no. 8, pp. 1788, 2019.

[8] F. A. Salaht, F. Desprez, and A. Lebre, "An Overview of Service Placement Problem in Fog and Edge Computing," in *ACM Comput. Surv*., vol. 53, issue. 3, no. 65, pp. 1-35, 2020.

[9] A. Yousefpour et al., "FOGPLAN: A Lightweight QoS-Aware Dynamic Fog Service Provisioning Framework," in *IEEE Internet of Things Journal*, vol. 6, no. 3, pp. 5080-5096, June 2019.

[10] B. Donassolo, I. Fajjari, A. Legrand and P. Mertikopoulos, "Fog Based Framework for IoT Service Provisioning," in *Proc. 2019 16th IEEE Annual Consumer Communications & Networking Conference (CCNC),* 2019, pp. 1-6,

[11] L. Mai, N. N. Dao, M. Park, "Real-Time Task Assignment Approach Leveraging Reinforcement Learning with Evolution Strategies for Long-Term Latency Minimization in Fog Computing", *Sensors*, vol. 18, no. 9, 2830, 2018.

[12] R. Mahmud, S. N. Srirama, K. Ramamohanarao, and R. Buyya, "Quality of experience (QoE)-aware placement of applications in Fog Computing environments," J. Parallel Distrib. Comput., vol. 132, pp. 190-203, 2019.

[13] O. Skarlat, S. Schulte, M. Borkowski, and P. Leitner, "Resource Provisioning for IoT Services in the Fog," in *9th IEEE International Conference on Service Oriented Computing and Applications (SOCA 2016). IEEE*, 2016, pp. 32–39.

[14] Q. T. Minh, D. T. Nguyen, A. Van Le, H. D. Nguyen and A. Truong, "Toward service placement on Fog computing landscape," 2017 *4th NAFOSTED Conference on Information and Computer Science*, 2017, pp. 291-296.

[15] M. Goudarzi, H. Wu, M.S. Palaniswami, R. Buyya, "An Application Placement Technique for Concurrent IoT Applications in Edge and Fog Computing Environments", in *IEEE Transactions on Mobile Computing*, vol. 20, no. 4, pp. 1298-1311, 1 April 2021.

[16] N. Auluck, O. Rana, S. Nepal, A. Jones and A. Singh, "Scheduling Real Time Security Aware tasks in Fog Networks," in *IEEE Transactions on Services Computing*, May 2019.

[17] G. R. Russo, V. Cardellini, F. L. Presti, M. Nardelli, "Towards a Security-Aware Deployment of Data Streaming Applications in Fog Computing," In: Chang W., Wu J. (eds) Fog/Edge Computing For Security, Privacy, and Applications. Advances in Information Security, vol 83. Springer, Cham., 2021.

[18] IBM CPLEX Optimizer, Available online: https://www.ibm.com/analytics/cplex-optimizer (Accessed: July 25, 2021)